\documentclass[12pt,a4paper]{article}
\usepackage{amssymb,amsmath,amsthm}
\usepackage{natbib}
\theoremstyle{plain}

\theoremstyle{definition}

\numberwithin{equation}{section}
\usepackage{fullpage}

\usepackage{cite}

\usepackage{graphicx}

\begin{document}

\title{Heterogeneous Beliefs with Partial Observations}\label{DBFilteringModel}

\author{ A.A. Brown \thanks{Wilberforce Road, Cambridge CB3 0WB, UK (phone = +44 1223 337969 , email = A.A.Brown@statslab.cam.ac.uk)} \thanks{I would like to thank Chris Rogers for his suggestion of this avenue of research and for many inciteful comments along the way.}  \\ \small Statistical Laboratory, \\  \small University of Cambridge  }

\date{First Version: April 2008  \\ This Version: July 2009 }

\maketitle

\abstract{This paper examines a heterogeneous beliefs model in which there is a process that is only partially observed by the agents. The economy contains a risky asset producing dividends continuously in time.   The dividends are observed by the agents.  The dividends are assumed to be a known function of some other unobserved process. The agents use filtering to estimate the value of this unobserved process. The agents have different beliefs about the dynamics of the unobserved process and therefore form different estimates.  We analyse this model and derive the state price density.  We use this to derive the riskless rate.  We also characterise the price of the risky asset in terms of the solution of a series of differential equations. }

\section{Introduction}

This paper will build on the theory outlined in \citet{BrownRogersDiverseBeliefs}. We consider an economy with unit supply of a risky asset and zero net supply of a riskless asset. The risky asset produces dividends continuously in time and these dividends are observed by all the agents.   The dividend process is assumed to be a function of another process, $X$, which is not observed by the agents.  In order for the agents to determine how to behave, they will use filtering to estimate the unobserved process.  Since the agents never observe this process, we will obtain a non-trivial steady state.

The model assumes that the dividend process is a linear function of the first component of the process $X$; this linear function is known to all the agents.   We will assume that $X$ is a multidimensional Ornstein-Uhlenbeck (OU) process. Since $X$ is a multidimensional OU process, all the components of $X$ will potentially affect the behaviour of the first component.  Agents can easily work out the first component of $X$ by looking at the dividend process, but they will be unable to work out the other components.  Since they need to know the other components to predict the behaviour of the dividend process, they will need to use filtering.

We will assume that the agents have different beliefs about the stochastic differential equation that governs $X$.  Specifically, the agents each have a different belief for the matrix that determines the dynamics of the OU process.  They will therefore each compute different estimates of $X$.  This will therefore affect how they behave.

The agents maximise the discounted expected utility of future consumption.  They each have CARA utility.  We can use the optimisation problem to characterise the equilibrium as in \citet{BrownRogersDiverseBeliefs}.  Using market clearing we may then determine the state price density.  We then use this to work out the interest rate process.  We can also explore the stock price, although we are unable to obtain a closed form expression.

There is a large literature on heterogeneous beliefs, which has been discussed in detail in \citet{BrownRogersDiverseBeliefs}.  Work includes \citet{Kurz}, \citet{Kurz94}, \citet{Kurz97}, \citet{KurzMotolese}, \citet{Kurz08}, \citet{KJM}, \citet{Fan}, \citet{HarrisonKreps}, \citet{Morris}, \citet{WuGuo03}, \citet{WuGuo04}, \citet{HarrisRaviv}, \citet{KandelPearson}, \citet{BuraschiJiltsov}, \citet{JouiniNapp}.  We will now briefly review some of the filtering models that are of a similar flavour to the one presented here.

\citet{XiongYan} consider a model in which there is a risky technology whose instantaneous return, $f_{t}$, obeys an OU process, but the mean of this OU process is not fixed.  This mean is unobserved by the agents and follows a different OU process.  Agents observe $f$ and another information process which is in fact independent of all the other processes in the system.  However, agents think the information process is correlated to the Brownian motion driving the unobserved OU process.  The two groups of agents have different beliefs about whether this correlation is positive or negative. The paper shows that trading volume increases with belief diversity.  The paper also looks at bond yields and uses the model to explain some effects that are observed in bond data.  \citet{DumasKurshevUppal} consider a similar model in which agents observe a dividend process which is obeying some SDE.  The agents do not know the drift of this SDE. This drift is random and obeys some other SDE.  Agents also observe an information signal.  One group of agents thinks that this signal is positively correlated with the dividend process.  The other group correctly thinks that it is not.  Agents must perform filtering to deal with the unknown drift. The authors derive asset prices and show that the presence of irrational agents leads to lower stock prices that are more volatile.

\citet{BrennanXia} consider a representative agent model in which the dividend process of a stock satisfies an SDE that has an unobserved drift.   The drift satisfies an OU process.  The agent maximises the expected discounted utility of future consumption.  The authors solve for the stock price and hence deduce quantities such as the equity premium and price dividend ratio.  They then calibrate the driving SDEs of their model to the Shiller data set and calculate various derived quantities.  Their results provide good agreement with the data, but they assume a constant of relative risk aversion of 15 and negative discounting, both of which are undesirable. 

\citet{ScheinkmanXiong} consider a model in which two groups of agents observe a dividend process that obeys an SDE with drift unknown to the agents.  This drift obeys some other SDE.  The agents observe two signals, called A and B.  Agents in group A think that the noise term in the SDE driving signal A is correlated to the noise term in the SDE driving the drift process.  Similarly, agents in group B think that the noise term in the SDE driving signal B is correlated to the noise term in the SDE driving the drift process.  In fact, the noise terms in the SDEs are independent.  Using standard filtering theory, the authors deduce the conditional means of each of the agents for the drift term.  They interpret this as the beliefs of the agents.  The authors derive equilibrium prices for the asset.  They then show that each agent is willing to pay more than their perceived fundamental value of the asset, since when they buy the asset, they are also buying an option to resell the asset at a possibly inflated price.  The authors use this to explain bubbles.

\citet{David}  introduces a model in which there are two fundamental processes in the economy.  The first is the output of the economy, of which each agent receives a fraction.  The second is a dividend process.  The drift of the dividend process and the economic output obey two SDEs.  The drifts of these SDEs are unobserved by the agents and follow a two-state continuous time Markov chain.  Agents form beliefs about the drifts based on their observations.  They then seek to maximise their expected utility.  David derives quantities such as the consumption and interest rate and uses his model to explain why the stock volatility can be so high.  \citet{DavidVeronesi}  also look at a continuous time model in which at any given time, the economy can be in one of two states; boom and recession.  The agents do not observe this state directly, but instead must infer it from their observations of the dividend process.

\citet{DetempleMurthy} consider a model with 3 different processes.  One process gives the total amount produced by the economy.  This process is observed.  However, its drift is unobserved and obeys a different SDE.  Finally, there is a third information process, observed by the agents.  The drift of this process is unobserved.  Agents work under different measures and form beliefs about the unknown drifts.  Agents maximise expected utility.  The paper derives an equilibrium for this model. 

The structure of this paper is as follows. Section \ref{DBFModel} explains the model in detail. Section  \ref{DBFDerivingSPD} then proceeds to analyse the model by deriving a state price density.  In section \ref{DBFUsingSPD} we look at the riskless rate and stock price. We can proceed far with the calculation of the stock price, but unfortunately are unable to obtain a closed form expression.  Finally, we conclude in section \ref{DBFConc}.

\section{The Model}\label{DBFModel}

We will now describe our model in detail.  As in \citet{BrownRogersDiverseBeliefs}, we assume that there is unit supply of a single risky asset.  This asset produces a dividend continuously in time. There is also a riskless asset in zero net supply.  The dividend of the risky asset at time $t$ is denoted $\delta_{t}$.  We will assume that $\delta_{t}$ is a function of the unobserved process $X_{t}$.  Agents disagree about the dynamics of $X$.  Thus, when they perform filtering they will form different estimates of $X$, which in turn will affect their behaviour.

We take our sample space to be $\Omega= C(\mathbb{R}_{+},\mathbb{R}^{n})$ and we let $X_{t}(\omega) \equiv \omega(t)$ denote the canonical process.  Let $\mathbb{P}_{0}$ denote the measure under which $X$ is a standard Brownian motion in $\mathbb{R}^{n}$.  

There are $J \ge 2$ agents in our model.  Each agent will work under his own measure.  Agent $j$ thinks that the true measure is given by $\mathbb{P}^{j}$, where $\mathbb{P}^{j}$ is defined by:
\begin{align*}
\Lambda_{t}^{j} := \frac{d\mathbb{P}^{j}}{d\mathbb{P}_{0}} = \exp \left[- \int_{0}^{t} (B^{j} X_{s}, dX_{s}) - \frac{1}{2} \int_{0}^{t} |B^{j} X_{s}|^{2} ds  \right] 
\end{align*}
where $B^{j}$ is an $N$x$N$  matrix\footnote{We will assume that all the eigenvalues of $B^{j}$ have negative real part.  This implies that under the measure $\mathbb{P}^{j}$, $X$ is stable in distribution (see Corollary 2.2 of \citet{BasakBhattacharya}).}.  The Cameron-Martin-Girsanov\footnote{See \citet{RogersWilliams}, IV.38 for an account} theorem gives that:
\begin{align*}
dW_{t}^{j}  := dX_{t} + B^{j} X_{t} dt   
\end{align*}
is a standard Brownian motion under $\mathbb{P}^{j}$.  Thus, $X$ is an OU process under $\mathbb{P}^{j}$.  

 None of the agents can observe $X$ directly, but they do observe the dividend process, $\delta_{t}$.  The dividend process is given by:
\begin{align*}
\delta_{t} =  a_{0} + \sigma e_{1} \cdot X_{t}
\end{align*}
for some constants $a_{0}$ and $\sigma$, which are known to the agents\footnote{Here, $e_{1}$ is the vector with 1 in the first position and zeros elsewhere}.  We assume that all the agents receive the $\emph{same}$ signal, namely the dividend process, $\delta_{t}$.  The important point is that in general $B^{i} \ne B^{j}$ for $i \ne j$, so their beliefs will differ.  For example, agent 1 may take $B^{1}$ to be diagonal and hence $\delta_{t}$ will behave as a one-dimensional OU process.  In contrast, agent 2 may construct $B^{2}$ in a way that means $\delta_{t}$ is affected by all the different components of $X_{t}$.  There are many different possible setups, but the exact form of the beliefs is not important at this stage.  

We define the observation filtration to be $\mathcal{Y}_{t}:= \sigma( e_{1} \cdot X_{s}: s \le t)$ .  The objective of the agent $j$ is:
\begin{align*}
\text{max} \quad \mathbb{E}^{j} \left[\int_{0}^{\infty} U_{j}(t,c^{j}_{t})  dt \right]
\end{align*}
where $c_{t}^{j}$ is the consumption of agent $j$ at time $t$ and $U_{j}(t, c_{t}^{j}) = -\frac{1}{\gamma_{j}} e^{-\gamma_{j} x} e^{-\rho t}$ is the discounted CARA utility.  Alternatively, we may write the objective as:
\begin{align}
\text{max} \quad \mathbb{E}^{0} \left[\int_{0}^{\infty}  U_{j}(t, c^{j}_{t}) \widehat{\Lambda}^{j}_{t}  dt \right]  \label{ReExpressedObj}
\end{align}
where $\widehat{\Lambda}^{j}_{t}$ is defined by:
\begin{align}
\widehat{\Lambda}^{j}_{t} := \mathbb{E}^{0}[\Lambda^{j}_{t} | \mathcal{Y}_{t} ]  \label{LambdaTildeDefn}
\end{align}
This completely specifies our model and we can now proceed to deduce a state price density.

\section{Deriving the State Price Density}\label{DBFDerivingSPD}
Now that we have the objective in the form (\ref{ReExpressedObj}), we are in the same case as in the paper of \citet{BrownRogersHBFiniteLivedAgents}.  Thus, we may proceed in exactly the same way to deduce the state price density.  We do this in two steps (see \citet{BrownRogersDiverseBeliefs} for more details).  Firstly, by looking at the price of an arbitrary contingent claim we can deduce that:   
\begin{align*}
\zeta_{t} \nu_{i} =  U_{i}'(t, c_{t}^{i}) \Lambda^{i}_{t}
\end{align*}
where $\nu_{i}$ is some some constant, different for each agent.   Taking logarithms and using market clearing, we may then deduce that:
\begin{align*}
\log \zeta_{t} = G - \rho t - \Gamma \sigma X_{t} + \frac{\Gamma}{J} \sum \frac{1}{\gamma_{j}} \log \widehat{\Lambda}^{j}_{t} 
\end{align*} 
where $\Gamma^{-1} = \sum_{j} \gamma_{j}^{-1}$ and $G$ is some constant.  We can use this state price density to deduce the interest rate and the stock price. 

\subsection{Calculating $\widehat{\Lambda}_{t}$}

In order to understand the state price density, we need to obtain an expression for $\widehat{\Lambda}^{j}_{t}$.  For the analysis that follows, we will omit the superscript $j$ unless doing so would cause confusion. 

First note that $\mathcal{Y}_{t}$ is generated by the process $X^{1}$ and furthermore that $X^{1}$ is a $\mathcal{Y}_{t}$-Brownian motion under $\mathbb{P}_{0}$.  Since $\widehat{\Lambda}_{t}$ is a $\mathcal{Y}_{t}$-martingale, we may therefore apply the martingale integral representation theorem\footnote{See \citet{RogersWilliams} IV.36} to deduce that:
\begin{align}
\widehat{\Lambda}_{t} = 1 + \int_{0}^{t} H_{s} dX_{s}^{1} \label{LamdaTildeBrownianIntegralRep}
\end{align}
for some $\mathcal{Y}_{t}$ process $H_{t}$.  

We will now determine the process $H$.  To do this, we introduce an arbitrary process $\theta_{t}$, which is assumed to be bounded and adapted to the filtration $(\mathcal{Y}_{t})_{t \ge0}$.  We have:
\begin{align*}
\mathbb{E}^{0}[ \widehat{\Lambda}_{t} \int_{0}^{t} \theta_{u} dX_{u}^{1} ] & = \mathbb{E}^{0}[ \Lambda_{t} \int_{0}^{t} \theta_{u} dX_{u}^{1} ] \\
& = \mathbb{E}^{0}[ \int_{0}^{t} \Lambda_{s} (-B X_{s}, dX_{s}) \int_{0}^{t} \theta_{u} dX_{u}^{1} ] \\
& = \mathbb{E}^{0}[ \int_{0}^{t} \Lambda_{u} \theta_{u} e_{1}' (-B X_{u}) du ] \\
& = \mathbb{E}^{0}[ \int_{0}^{t} \theta_{u} e_{1}' \widehat{(- \Lambda_{u} B X_{u})} du ] 
\end{align*}
where $\hat{Y}_{u}$ denotes $\mathbb{E}^{0}[Y_{u}| \mathcal{Y}_{u}]$ and $e_{i}$ denotes the column-vector with 1 in the $i$-th position and zeroes elsewhere.  Here, we have used the definition of conditional expectation in the first and final lines, the integral representation of $\Lambda$ in the second line and the fact that $\mathbb{E} AB = \mathbb{E} [A,B]$ in the third line.  However, we may also use (\ref{LamdaTildeBrownianIntegralRep}) to deduce that the above expression is:
\begin{align*}
\mathbb{E}^{0}[ \int_{0}^{t} H_{u} \theta_{u} du ]
\end{align*}
Thus we obtain:
\begin{align*}
\mathbb{E}^{0}\left[ \int_{0}^{t}  \theta_{u} \big( H_{u} + e_{1}' \widehat{(\Lambda_{u} B X_{u})} \big)du \right] =0
\end{align*}
But since $(\theta_{t})_{t \ge 0}$ is just an arbitrary $(\mathcal{Y}_{t})_{t \ge 0}$-process, we deduce that 
\begin{align*}
H_{u} &=  -e_{1}' \widehat{(\Lambda_{u} B X_{u})} \\
 & = \frac{\mathbb{E}^{0}[-\Lambda_{u} e_{1}' B X_{u}  |\mathcal{Y}_{u}] \widehat{\Lambda}_{u} }{\mathbb{E}^{0}[\Lambda_{u} | \mathcal{Y}_{u}]} \\
& = \mathbb{E}[ -e_{1}' B X_{u}  |\mathcal{Y}_{u}] \widehat{\Lambda}_{u} \\
& = - e_{1}' B \widehat{X}_{u} \widehat{\Lambda}_{u}
\end{align*}
where $\mathbb{E}$ denotes expectation under the agent's measure.  Combining this with (\ref{LamdaTildeBrownianIntegralRep}), we have:
\begin{align}
d\widehat{\Lambda}_{u} = - \widehat{\Lambda}_{u} e_{1}' B \widehat{X}_{u} dX_{u}^{1}  \label{LamdaTildeSDE} 
\end{align}
We therefore need to develop $\widehat{X}$ in order to better understand $\widehat{\Lambda}$.

\subsection{Deriving $\widehat{X}_{u}$}

First define:
\begin{align*}
h_{t}^{i} = - e_{i}' (B X_{t})
\end{align*}
Then we have that:
\begin{align*}
dX_{t}^{1} = dW^{1}_{t} + h^{1}_{t} dt
\end{align*}
Also define:
\begin{align}
N_{t} = X^{1}_{t} - \int_{0}^{t} \widehat{h^{1}_{s}} ds  \label{DBFMNDefn}
\end{align}
Furthermore, define the conditional covariance matrix $V$ by:
\begin{align*}
V^{ij}_{t} = \widehat{X_{t}^{i} X_{t}^{j}} - \widehat{X_{t}^{i}}\widehat{X_{t}^{j}}
\end{align*}
We wish to derive the distribution of $\widehat{X}_{t}$.  Note that since $(X_{t}^{1},X_{t})$ is a Gaussian process, the conditional law of $X_{t}$ given $\mathcal{Y}_{t}$ will be Gaussian.  Thus, we just need to determine the conditional mean and conditional covariance matrix. It is straightforward to deduce\footnote{See the appendix} that:
\begin{align}
\boxed{\widehat{X_{t}^{i}} = \widehat{X_{0}^{i}} - \int_{0}^{t} ( B \widehat{X_{s}})_{i} ds - \int_{0}^{t} \left( (BV)_{1i} - \delta_{i1}   \right) dN_{s}} \label{EqnForXhat}
\end{align}
Alternatively, in vector notation:
\begin{align*}
\widehat{X_{t}} = \widehat{X_{0}} - \int_{0}^{t} ( B \widehat{X_{s}}) ds - \int_{0}^{t} \left( ((BV)')_{\cdot 1} - e_{1}   \right) dN_{s}
\end{align*}
We may also deduce that:
\begin{align}
\boxed{dV_{s}^{ij} = - \left[(BV)_{ji} + (BV)_{ij} - \delta_{ij} + ((BV)_{1i} - \delta_{i1})((BV)_{1j} - \delta_{j1})   \right] ds} \label{DBFMEqnForV}
\end{align}
Note that there is no $dN$ term in the above SDE and thus $V$ is in fact deterministic.  We may also use (\ref{DBFMEqnForV}) to specialise to the case in which $\dot{V}=0$, but we first introduce some notation.

\subsubsection{Notation}

Since $V^{1j}_{t}=V^{i1}_{t}=0$, we may write:
\begin{align*}
V_{t}= \begin{pmatrix}
0 & 0_{n-1}'   \\
0_{n-1} &  \tilde{V}_{t}  
\end{pmatrix}
\end{align*}
where $0_{n-1}$ denotes the zero-vector of length $n-1$.  Similarly, we may write:
\begin{align*}
B = \begin{pmatrix}
b_{11} & C'   \\
A &  \tilde{B}  
\end{pmatrix}
\end{align*}
Furthermore, we may define $x_{t}$ and $\widehat{x_{t}}$ by $\widehat{X_{t}} = (x_{t}, \widehat{x_{t}})'$.  We are of course only interested in working out $\widehat{x_{t}}$.  From (\ref{EqnForXhat}) we deduce that:
\begin{align*}
d\widehat{x_{t}} = - (A x_{t} + \tilde{B} \widehat{x_{t}}) dt - (\tilde{V} C) dN_{t}
\end{align*}
But, 
\begin{align*}
dN_{t} = dx_{t} + (b_{11} x_{t} + C' \widehat{x_{t}}) dt
\end{align*}
so we learn that:
\begin{align}
d\widehat{x_{t}}= -(A+ \tilde{V}C b_{11}) x_{t} dt - \tilde{V} C dx_{t} - (\tilde{B} + \tilde{V} C C' ) \widehat{x_{t}} dt  \label{xhateqn}
\end{align}

\subsubsection{Stationarity}

We now can examine the steady state in which $V$ is not changing.  We want to solve $\dot{V}=0$, or equivalently:
\begin{align}
\left[(BV)_{ji} + (BV)_{ij} - \delta_{ij} + ((BV)_{1i} - \delta_{i1})((BV)_{1j} - \delta_{j1})   \right]=0  \label{StationaryEqnForV}
\end{align}
This equation is automatically satisfied whenever $i=1$ or $j=1$. For $i,j>1$, we have:
\begin{align*}
(BV)_{ij}= \sum_{k=1}^{n} B_{ik}V_{kj} = \sum_{k=2}^{n} \tilde{B}_{i(k-1)} \tilde{V}_{(k-1)j} = \tilde{B}\tilde{V}_{ij}
\end{align*}
and:
\begin{align*}
BV_{1i}= \sum_{k} B_{1k}V_{ki} = \sum_{k=1}^{n-1} C_{k} \tilde{V}_{ki} = (C' \tilde{V})_{i}
\end{align*}
Substituting in (\ref{StationaryEqnForV}) we see that the stationary covariance matrix is specified by:
\begin{align*}
\boxed{\tilde{B} \tilde{V} +  \tilde{V}' \tilde{B}' + \tilde{V}' C C' \tilde{V} = I_{n-1}}
\end{align*}

\subsection{The State Price Density}

\subsubsection{Notation for multiple agents}

Now that we have developed our expressions for $\widehat{\Lambda}$ and $\widehat{X}$, we may use these to shed light on the state price density.  

Before we do this, we define some further notation to deal with the $J$ different agents.  Recall that each agent has a different $B$ matrix, and hence a different $\tilde{B}, C$ and $A$ matrix. Let $\hat{x}^{j}_{t}$ be the estimate of $(X_{t}^{2}, X_{t}^{3}, ..., X_{t}^{n})'$ according to agent $j$.  Then we may define: 
\begin{align*}
Z_{t} = (\hat{x}_{t}^{1}, \hat{x}_{t}^{2}, ..., \hat{x}_{t}^{J})'
\end{align*}
to be the concatenation of the estimates of all the different agents of the unknown components of $X$.  Using equation (\ref{xhateqn}) we see that $Z_{t}$ satisfies the SDE:
\begin{align*}
dZ_{t} = (A_{1} x_{t} + B_{1} Z_{t}) dt + Q_{1} dx_{t}
\end{align*}
Here, $A_{1}$ is the stacked version of $-(A+\tilde{V}C b_{11})$ for each agent and $Q_{1}$ is a stacked version of $- \tilde{V} C$ for all the different agents (recall that these matrices are different for each agent).  $B_{1}$ is the $((n-1)J)$x$((n-1)J)$ block diagonal matrix with each of the $-(\tilde{B} + \tilde{V} C C')$ of each of the different agents in the $J$ blocks.  

We also define $\bar{Z}_{t}$ = $(x_{t}, Z_{t})'$.  Then we have that:
\begin{align*}
d\bar{Z}_{t} = \bar{B} \bar{Z}_{t} dt + \bar{Q} dx_{t}
\end{align*} 
where $\bar{Q} = (1, Q_{1}')'$ and 
\begin{align*}
\bar{B} = \begin{pmatrix} 0 & 0_{J(n-1)} \\
A_{1} & B_{1} \end{pmatrix}
\end{align*}

\subsubsection{Final expression for the state price density}

Having defined this notation, we may now use it to derive our final expression for the state price density.  

We know from (\ref{LamdaTildeSDE}) that the $\widehat{\Lambda^{j}}$ satisfies the SDE:   
\begin{align*}
d\widehat{\Lambda^{j}_{u}} = - \widehat{\Lambda^{j}_{u}} e_{1}' B^{j} \widehat{X^{j}_{u}} dX_{u}^{1} 
\end{align*}
The solution to this SDE is given by:
\begin{align*}
 \widehat{\Lambda^{j}_{T}} = \widehat{\Lambda^{j}_{0}} \exp \left[\int_{0}^{T} - (e_{1}' B^{j}) \hat{X}^{j}_{u} dx_{u} - \frac{1}{2} \int_{0}^{T} (e_{1}' B^{j} \hat{X}^{j}_{u})^{2} du  \right]
\end{align*}
In view of the expression we have for the state price density, we will need:
\begin{align*}
 \frac{1}{\gamma_{j}} \log \widehat{\Lambda^{j}_{T}} & = \frac{1}{\gamma_{j}} \log \widehat{\Lambda^{j}_{0}} + \int_{0}^{T} - \frac{(e_{1}' B^{j})}{\gamma_{j}} \hat{X}^{j}_{u} dx_{u} - \frac{1}{2} \int_{0}^{T} \frac{(e_{1}' B^{j} \hat{X}^{j}_{u})^{2}}{\gamma_{j}} du   \\
& =  \frac{1}{\gamma_{j}} \log \widehat{\Lambda^{j}_{0}} + \int_{0}^{T} \big(- \frac{b_{11}^{j}}{\gamma_{j}} x_{u} - \frac{(C^{j})'}{\gamma_{j}} \hat{x}^{j}_{u} \big) dx_{u} - \frac{1}{2} \int_{0}^{T} \gamma_{j} \big(\frac{b_{11}^{j}}{\gamma_{j}} x_{u} + \frac{(C^{j})'}{\gamma_{j}} \hat{x}^{j}_{u} \big)^{2}  du  
\end{align*}
Specifically, we are interested in the sum of these terms, which can be expressed as:
\begin{align*}
\sum_{j=1}^{J} \frac{1}{\gamma_{j}} \log \widehat{\Lambda^{j}_{T}} & = const. + \int_{0}^{T} \big(- \sum_{j} \frac{b_{11}^{j}}{\gamma_{j}} x_{u} - \sum_{j} \frac{(C^{j})'}{\gamma_{j}} \hat{x}^{j}_{u} \big) dx_{u} \\
& - \frac{1}{2} \int_{0}^{T} \big(\sum_{j} \frac{(b_{11}^{j})^{2}}{\gamma_{j}} x_{u}^{2} + 2 \sum_{j} \frac{b_{11}^{j} (C^{j})'}{\gamma_{j}} x_{u} \hat{x}^{j}_{u} + \sum_{j} \frac{(\hat{x}^{j}_{u})' C^{j} (C^{j})'\hat{x}^{j}_{u}}{\gamma_{j}} \big)  du  \\
& = const. + \int_{0}^{T} \bar{\alpha}' \bar{Z_{t}} dx_{t} - \frac{1}{2} \int_{0}^{T} \bar{Z_{t}}' \bar{\beta} \bar{Z_{t}} dt 
\end{align*}
where $\bar{\alpha}$ is a $1+J(n-1)$ dimensional vector given by:
\begin{align*}
\bar{\alpha} = - ( \sum_{j} \frac{b_{11}^{j}}{\gamma_{j}}, \frac{(C^{1})'}{\gamma_{1}}, \frac{(C^{2})'}{\gamma_{2}}, ... , \frac{(C^{J})'}{\gamma_{J}})'
\end{align*}
and
\begin{align*}
\bar{\beta} = \begin{pmatrix} \sum_{j} \frac{(b_{11}^{j})^{2}}{\gamma_{j}} & \frac{b_{11}^{1} (C^{1})'}{\gamma_{1}} & \frac{b_{11}^{2} (C^{2})'}{\gamma_{2}} & ... & \frac{b_{11}^{J} (C^{J})'}{\gamma_{J}} \\
\frac{b_{11}^{1} C^{1}}{\gamma_{1}} & \frac{C^{1} (C^{1})'}{\gamma_{1}} & 0_{n-1,n-1} & ... & 0_{n-1,n-1} \\
\frac{b_{11}^{2} C^{2}}{\gamma_{2}} &  0_{n-1,n-1} & \frac{C^{2} (C^{2})'}{\gamma_{2}}  & ... & 0_{n-1,n-1} \\
\vdots & \vdots & & \ddots & \vdots \\
\frac{b_{11}^{J} C^{J}}{\gamma_{J}} &  0_{n-1,n-1} & 0_{n-1,n-1} & ... & \frac{C^{J} (C^{J})'}{\gamma_{J}}  \end{pmatrix}
\end{align*}
where $0_{n-1,n-1}$ denotes the zero matrix of dimension $(n-1)$x$(n-1)$.  Note that in the above notation the first row and column have dimension 1, whereas all subsequent rows and columns have dimension $n-1$.

The final expression we obtain for the state price density is given by:
\begin{align*}
\boxed{\log \zeta_{T} = const. - \rho T - \Gamma \sigma x_{T} + \frac{\Gamma}{J} \big( \int_{0}^{T} \bar{\alpha}' \bar{Z_{t}} dx_{t} - \frac{1}{2} \int_{0}^{T} \bar{Z_{t}}' \bar{\beta} \bar{Z_{t}} dt \big)}
\end{align*}

\section{Using the State Price Density}\label{DBFUsingSPD}

We now have a very explicit form for the state price density, which can be used to price any assets.  We will derive the riskless rate in this model and also illustrate the method for calculating the stock price.

\subsection{Riskless rate}

Recall that there is a single riskless asset in the model, which is in zero net supply.  The interest rate must satisfy:
\begin{align*}
d \zeta_{t} = \zeta_{t} ( - r_{t} dt - \kappa_{t} dx_{t})
\end{align*}
where $(r_{t})_{t \ge 0}$ is the interest rate process and $(\kappa_{t})_{t \ge 0}$ is some other process, not currently of interest to us.  Applying It\^{o}'s formula to our expression for the state price density, we may simply read off the interest rate process as:
\begin{align*}
r_{t} = \rho + \dfrac{\Gamma}{2 J} \bar{Z}_{t}' \bar{\beta} \bar{Z}_{t} - \dfrac{\Gamma}{2} ( \dfrac{1}{J} \bar{\alpha}' \bar{Z}_{t} - \sigma)^{2}
\end{align*}

\subsubsection{Remarks on the riskless rate}

We see that if the agents are impatient ($\rho$ large) then the riskless rate offered to them must be higher to stop them simply consuming their wealth.  If we consider the case in which $\bar{Z}_{t}=0$, we see that the riskless rate is simply $\rho -\dfrac{\Gamma}{2} \sigma^{2}$, so that a larger volatility in the dividend process means that the riskless rate is lower.   The dependence on $\bar{Z}$ is more complex, since it will in turn depend on our assumptions for $\bar{\alpha}$ and $\bar{\beta}$.
 
\subsection{The Stock Price}

The stock price is given by:
\begin{align*}
S_{t} = \mathbb{E}_{t}^{0} \big[\int_{t}^{\infty} \frac{\zeta_{T} \delta_{T}}{\zeta_{t}} dT \big]
\end{align*}
\begin{multline*}
= \int_{t}^{\infty} e^{-\rho (T-t)} \mathbb{E}_{t}^{0} \big[ \exp \big(-\Gamma \sigma (x_{T}-x_{t}) \\
+ \frac{\Gamma}{J} \big( \int_{t}^{T} \bar{\alpha}' \bar{Z_{u}} dx_{u} - \frac{1}{2} \int_{t}^{T} \bar{Z_{u}}' \bar{\beta} \bar{Z_{u}} du \big) \big) (\sigma x_{T} + a_{0}) \big] dT
\end{multline*}
If we can calculate:
\begin{align*}
\mathbb{E}_{t}^{0} \big[ \exp \big(-\Gamma \sigma (x_{T}-x_{t}) + \frac{\Gamma}{J} \big( \int_{t}^{T} \bar{\alpha}' \bar{Z_{u}} dx_{u} - \frac{1}{2} \int_{t}^{T} \bar{Z_{u}}' \bar{\beta} \bar{Z_{u}} du \big) + \theta (\sigma x_{T} + a_{0}) \big) \big]
\end{align*}
then we may differentiate with respect to $\theta$ to obtain the conditional expectation that we require.

\subsubsection{Calculating the conditional expectation}

We will calculate:
\begin{multline}
V^{T}(t, \bar{Z}_{t}; \theta) : = \mathbb{E}_{t}^{0} \big[ \exp \{-\Gamma \sigma (x_{T}-x_{t}) + \frac{\Gamma}{J} \big( \int_{t}^{T} \bar{\alpha}' \bar{Z_{u}} dx_{u} \\
- \frac{1}{2} \int_{t}^{T} \bar{Z_{u}}' \bar{\beta} \bar{Z_{u}} du \big) + \theta (\sigma x_{T} + a_{0}) \} \big]  \label{Vdefn}
\end{multline}
We will show that:
\begin{align*}
V^{T}(t,\bar{Z}_{t}; \theta) = \exp \big(\frac{1}{2} (\bar{Z}_{t})' a(\tau) \bar{Z}_{t} + b(\tau) \bar{Z}_{t} + c(\tau)  \big)
\end{align*}
where $\tau = T-t$ and $a(\tau)$ is a symmetric $(J(n-1)+1)$x$(J(n-1)+1)$ matrix, $b(\tau)$ is a $(J(n-1)+1)$ row vector and $c(\tau)$ is a scalar. We will omit the explicit dependence on $\tau$ except where we specifically require it.  
In order to calculate $a,b$ and $c$, we will use a martingale argument.  We define:
\begin{align*}
M^{T}_{t} & = \mathbb{E}_{t}^{0} \big[ \exp \{-\Gamma \sigma x_{T} + \frac{\Gamma}{J} \big( \int_{0}^{T} \bar{\alpha}' \bar{Z_{u}} dx_{u} - \frac{1}{2} \int_{0}^{T} \bar{Z_{u}}' \bar{\beta} \bar{Z_{u}} du \big) + \theta (\sigma x_{T} + a_{0}) \} \big] \\
& = V^{T}(t,\bar{Z}_{t}; \theta) \exp \Big[-\Gamma \sigma x_{t} + \frac{\Gamma}{J} \big(\int_{0}^{t} \bar{\alpha}' \bar{Z_{u}} dx_{u} - \frac{1}{2} \int_{0}^{t} \bar{Z_{u}}' \bar{\beta} \bar{Z_{u}} du  \big) \Big]
\end{align*}  
Since $(M_{t}^{T})_{0 \le t \le T}$ is a martingale we will apply It\^{o}'s formula to it and deduce that the term in $dt$ is zero.  This will enable us to get a series of three differential equations for $a,b$ and $c$.  Firstly, we have that:
\begin{multline*}
d( \frac{1}{2} \bar{Z}_{t}' a \bar{Z}_{t} + b \bar{Z}_{t} + c ) = \big(-\frac{1}{2} \bar{Z}_{t}' \dot{a} \bar{Z}_{t} - \dot{b} \bar{Z}_{t} - \dot{c} + b \bar{B} \bar{Z}_{t} +  \bar{Z}_{t}' a \bar{B} \bar{Z}_{t} + \frac{1}{2} \bar{Q}' a \bar{Q} \big) dt \\
+ \big(b \bar{Q} + \bar{Z}' a \bar{Q}  \big) dx_{t}
\end{multline*} 
Thus, 
\begin{multline*}
\frac{dV}{V} = \big(-\frac{1}{2} \bar{Z}_{t}' \dot{a} \bar{Z}_{t} - \dot{b} \bar{Z}_{t} - \dot{c} + b \bar{B} \bar{Z}_{t} +  \bar{Z}_{t}' a \bar{B} \bar{Z}_{t} + \frac{1}{2} \bar{Q}' a \bar{Q} + \frac{1}{2}(b \bar{Q} + \bar{Z}' a \bar{Q})^{2} \big) dt \\
+ \big(b \bar{Q} + \bar{Z}' a \bar{Q}  \big) dx_{t}  
\end{multline*}
Now define:
\begin{align*}
K(t, \bar{Z}_{t}) = - \Gamma \sigma x_{t} + \frac{\Gamma}{J} \big(\int_{0}^{t} \bar{\alpha}' \bar{Z_{u}} dx_{u} - \frac{1}{2} \int_{0}^{t} \bar{Z}_{u}' \bar{\beta} \bar{Z}_{u} du  \big) 
\end{align*}
Then, we have 
\begin{align*}
\frac{d( \exp\{K(t, \bar{Z}_{t})\})}{\exp\{K(t, \bar{Z}_{t})\}} = \Gamma ( \frac{\bar{\alpha}' \bar{Z}_{t}}{J} - \sigma ) dx_{t} + \Big(- \frac{\Gamma}{2 J} \bar{Z}_{t}' \bar{\beta} \bar{Z_{t}} + \frac{1}{2} \Gamma^{2} \big(\frac{\bar{\alpha}'\bar{Z}_{t}}{J} - \sigma  \big)^{2}  \Big)dt 
\end{align*}
So finally we may calculate $dM$:
\begin{multline*}
\frac{dM_{t}^{T}}{M_{t}^{T}} = \big(-\frac{1}{2} \bar{Z}_{t}' \dot{a} \bar{Z}_{t} - \dot{b} \bar{Z}_{t} - \dot{c} + b \bar{B} \bar{Z}_{t} +  \bar{Z}_{t}' a \bar{B} \bar{Z}_{t} + \frac{1}{2} \bar{Q}' a \bar{Q} +\frac{1}{2} (b \bar{Q} + \bar{Z}' a \bar{Q})^{2} \big) dt \\
+\Big(- \frac{\Gamma}{2 J} \bar{Z}_{t}' \bar{\beta} \bar{Z_{t}} + \frac{1}{2} \Gamma^{2} \big(\frac{\bar{\alpha}'\bar{Z}_{t}}{J} - \sigma  \big)^{2}  \Big) dt + \Gamma ( \frac{\bar{\alpha}' \bar{Z}_{t}}{J} - \sigma ) \big(b \bar{Q} + \bar{Z}' a \bar{Q}  \big) dt  + \{ ... \} dx_{t}
\end{multline*}
Since $M$ is a martingale, the $dt$ term in the above equation must be zero.  Thus we obtain:
\begin{multline}
\frac{1}{2}\bar{Z}_{t}' \{ -\dot{a} + 2 a \bar{B}  - \frac{\Gamma}{ J} \bar{\beta} +  a \bar{Q} \bar{Q}' a + \frac{\Gamma^{2}}{ J^{2}} \bar{\alpha} \bar{\alpha}' + \frac{2 \Gamma a \bar{Q} \bar{\alpha}'}{ J}  \} \bar{Z}_{t} \\
+ \{-\dot{b} + b \bar{B} +  b \bar{Q} \bar{Q}' a - \frac{\Gamma^{2} \sigma \bar{\alpha}'}{J} + \frac{\Gamma b \bar{Q} \bar{\alpha}'}{J} - \Gamma \sigma \bar{Q}' a  \} \bar{Z}_{t} \\
+\{ -\dot{c} + \frac{1}{2} \bar{Q}' a \bar{Q} + \frac{1}{2}b \bar{Q} \bar{Q}' b' + \frac{1}{2} \Gamma^{2} \sigma^{2} - \Gamma \sigma b \bar{Q} \} = 0  \label{dMmgCond}
\end{multline}
Each of the above expressions in the parentheses must be equal to zero.  We also have some boundary conditions; note from (\ref{Vdefn}) that:
\begin{align*}
V^{T}(T, \bar{Z}_{T}; \theta) = \exp \left(\theta \sigma x_{T} + \theta a_{0}  \right)
\end{align*}
Thus, we have the boundary conditions:
\begin{align*}
a(0)= 0_{J(n-1), J(n-1)} \qquad b(0) = (\theta \sigma, 0_{J(n-1)})' \qquad c(0) = \theta a_{0}
\end{align*}
Unfortunately, the first line of (\ref{dMmgCond}) gives us a matrix Riccati equation.  It seems that solving this equation is intractable, and thus we are unable to proceed further with the stock price.  

\subsubsection{A PDE approach}

An alternative method for tackling the stock price is to use a PDE approach.  We may argue that $\zeta_{t} S_{t} + \int_{0}^{t} \zeta_{s} \delta_{s} ds$ is a martingale.  We can then apply It\^{o}'s formula and deduce that the $dt$ term is zero.  This will give us a PDE for the stock price. 

Proceeding in this way, first define $S_{t} = h(\bar{Z}_{t})$.  Then we may deduce that the stock price must satisfy the PDE:
\begin{align*}
 \triangledown h(\bar{Z}_{t}) \cdot  (\bar{B} \bar{Z}) + \dfrac{1}{2} \bar{Q}' H(\bar{Z}_{t}) \bar{Q} + ( \dfrac{\Gamma}{J} \bar{\alpha}' \bar{Z} - \Gamma \sigma ) \triangledown h(\bar{Z}_{t}) \cdot \bar{Q}   - r_{t} h(\bar{Z}_{t}) + \delta_{t}  = 0
\end{align*} 
where $H$ denotes the Hessian of $h$.  Unfortunately, there seems to be little hope of solving this PDE, given that the dimension of this problem is $J(n-1)+1$.

\section{Conclusions}\label{DBFConc}

We have introduced a very general model for dealing with heterogeneous beliefs of agents.  This model assumes that there is some unobserved process $X$ that drives the dividend process.  Agents differ in their views about the SDE that $X$ obeys.  This affects their behaviour.  We are able to proceed far with the analysis of this model, in particular, deriving the riskless rate in this model.  However, the calculation of the stock price appears intractable because it requires a solution of a matrix Riccati equation.

\bibliography{references}
\bibliographystyle{jmb}

\newpage

\appendix
\begin{center}
  {\bf APPENDIX}
\end{center}

\section{Deriving expressions (\ref{EqnForXhat}) and (\ref{DBFMEqnForV})}

We seek to determine an SDE for $\widehat{X}$. In order to do this, we will follow section VI.8-9 of \citet{RogersWilliams}.

First recall that $N$ was defined in (\ref{DBFMNDefn}) by:
\begin{align*}
N_{t} = X^{1}_{t} - \int_{0}^{t} \widehat{h^{1}_{s}} ds
\end{align*}
By Theorem VI.8.4 of \citet{RogersWilliams} we know that $N_{t}$ is a $\mathcal{Y}_{t}$ Brownian motion.  

Let $f: \mathcal{R}^{n} \mapsto \mathcal{R}$ and let $f_{t} := f(X_{t})$. Let $\mathcal{G}f_{t}$ be the function such that:
\begin{align*}
M_{t} = f_{t} - f_{0} - \int_{0}^{t} \mathcal{G}f_{s} ds
\end{align*}
is an $\mathcal{F}_{t}$-martingale.  Further, let $\alpha_{t}$ be such that:
\begin{align*}
df_{t} dX_{t}^{1} = \alpha_{t} dt
\end{align*}
Then by (VI.8.16) of \citet{RogersWilliams}, we have:
\begin{align}
\widehat{f}_{t} = \widehat{f_{0}} + \int_{0}^{t} \widehat{\mathcal{G} f_{s}} ds + \int_{0}^{t} \left(\widehat{f_{s}h_{s}^{1}} - \widehat{f_{s}}\widehat{h^{1}_{s}} + \widehat{\alpha_{s}} \right) dN_{s} \label{RandW816}
\end{align}
We will use (\ref{RandW816}) to derive the distribution of $\widehat{X}_{t}$.  Note that since $(X_{t}^{1},X_{t})$ is a Gaussian process, the conditional law of $X_{t}$ given $\mathcal{Y}_{t}$ will be Gaussian.  Thus, we just need to determine the conditional mean and conditional covariance matrix.  We do this by applying (\ref{RandW816}) to the functions:
\begin{align*}
f^{i}: x \mapsto x_{i} \\
f^{ij}: x \mapsto x_{i} x_{j}
\end{align*}  
Firstly, working on the $f^{i}$, we have:
\begin{align*}
dX_{t}^{i} = dW_{t}^{i} + h_{t}^{i} dt
\end{align*}
Noting that $df_{t}^{i} dX_{t}^{1} = dW_{t}^{i} dW_{t}^{1} = \delta_{i1} dt$, we have that:
\begin{align*}
\mathcal{G} f^{i}_{t} = h_{t}^{i} \qquad \alpha^{i} = \delta_{i1}
\end{align*}
If we put all this in (\ref{RandW816}), we obtain:
\begin{align*}
\widehat{X_{t}^{i}} = \widehat{X_{0}^{i}} - \int_{0}^{t} e_{i}' B \widehat{X_{s}} ds - \int_{0}^{t} e_{1}' \left( \widehat{X_{s}^{i}  B X_{s}} - \widehat{X_{s}^{i}} e_{1}' B \widehat{X_{s}} - \delta_{i1}   \right) dN_{s}
\end{align*}
Upon substituting for $V$, we obtain (\ref{EqnForXhat}).

We now move onto $f^{ij}$.  First note that:
\begin{align*}
d(X_{t}^{i}X_{t}^{j}) = X_{t}^{i} dW_{t}^{j} + X_{t}^{j} dW_{t}^{i} + [ X_{t}^{i} h_{t}^{j} + X_{t}^{j} h_{t}^{i} + \delta_{ij}] dt 
\end{align*}
and that:
\begin{align*}
df_{t}^{ij} dX_{t}^{1} = X_{t}^{i} \delta_{j1} dt + X_{t}^{j} \delta_{i1} dt
\end{align*}
Thus, we obtain:
\begin{align*}
\mathcal{G} f_{t}^{ij} = X_{t}^{i} h_{t}^{j} + X_{t}^{j} h_{t}^{i} + \delta_{ij} \qquad \alpha^{ij}_{t} = X_{t}^{i} \delta_{j1}  + X_{t}^{j} \delta_{i1} 
\end{align*}
Applying (\ref{RandW816}), we obtain:
\begin{multline*}
\widehat{X_{t}^{i}X_{t}^{j}} = \widehat{X_{0}^{i}X_{0}^{j}} + \int_{0}^{t} \left (\widehat{X_{s}^{i} h_{s}^{j}} + \widehat{X_{s}^{j} h_{s}^{i}} + \delta_{ij}  \right) ds \\
+ \int_{0}^{t} \left[\widehat{X_{s}^{i} X_{s}^{j} h_{s}^{1}} - \widehat{X_{s}^{i} X_{s}^{j}}\widehat{h_{s}^{1}} + \widehat{X_{s}^{i}} \delta_{j1} + \widehat{X_{s}^{j}} \delta_{i1}  \right] dN_{s} 
\end{multline*}
\begin{multline*}
=  \widehat{X_{0}^{i}X_{0}^{j}} - \int_{0}^{t} \left ((B \widehat{ X_{s})_{j} X_{s}^{i} } + (B\widehat{ X_{s})_{i}X_{s}^{j}} - \delta_{ij}  \right) ds \\ 
- \int_{0}^{t} \left[\sum_{k} B_{1k}\widehat{X_{s}^{k} X_{s}^{i} X_{s}^{j}} - \sum_{k} B_{1k} \widehat{X_{s}^{k}}\widehat{X_{s}^{i} X_{s}^{j}} - \widehat{X_{s}^{i}} \delta_{j1} - \widehat{X_{s}^{j}} \delta_{i1}  \right] dN_{s}
\end{multline*}
We need to work out terms of the form $\widehat{X_{s}^{i}X_{s}^{j}X_{s}^{k}}$.  However, since $X$ is a Gaussian process, we have that:
\begin{align*}
\widehat{X_{s}^{i}X_{s}^{j}X_{s}^{k}} = \widehat{X_{s}^{i}}V_{s}^{jk} + \widehat{X_{s}^{j}}V_{s}^{ik} + \widehat{X_{s}^{k}}V_{s}^{ij} + \widehat{X_{s}^{i}}\widehat{X_{s}^{j}}\widehat{X_{s}^{k}}
\end{align*}
If we now define $M_{t}^{ij} := \widehat{X_{t}^{i}} \widehat{X_{t}^{j}}$, we obtain:
\begin{multline*}
\widehat{X_{t}^{i}X_{t}^{j}}=  \widehat{X_{0}^{i}X_{0}^{j}} - \int_{0}^{t} \left (\sum_{k} B_{jk} \widehat{X_{s}^{k} X_{s}^{i} } + \sum_{k} B_{ik} \widehat{ X_{s}^{k}X_{s}^{j}} - \delta_{ij}  \right) ds \\ 
- \int_{0}^{t} \Big[\sum_{k} B_{1k}\widehat{X_{s}^{i}}V_{s}^{jk} + \sum_{k} B_{1k}\widehat{X_{s}^{j}}V_{s}^{ik} + \sum_{k} B_{1k}\widehat{X_{s}^{k}}V_{s}^{ij} \\
+\sum_{k} B_{1k}\widehat{X_{s}^{i}}\widehat{X_{s}^{j}}\widehat{X_{s}^{k}}  - \sum_{k} B_{1k} \widehat{X_{s}^{k}}\widehat{X_{s}^{i} X_{s}^{j}} - \widehat{X_{s}^{i}} \delta_{j1} - \widehat{X_{s}^{j}} \delta_{i1}  \Big] dN_{s}
\end{multline*}
\begin{multline*}
=  \widehat{X_{0}^{i}X_{0}^{j}} - \int_{0}^{t} \left (\sum_{k} B_{jk} (V+M)_{ki} + \sum_{k} B_{ik} (V+M)_{kj} - \delta_{ij}  \right) ds \\ 
- \int_{0}^{t} \Big[ (BV)_{1j}\widehat{X_{s}^{i}} + (BV)_{1i}\widehat{X_{s}^{j}} + (B\widehat{X_{s}})_{1} V_{s}^{ij} \\
-(B \widehat{X_{s}})_{1} V_{s}^{ij} - \widehat{X_{s}^{i}} \delta_{j1} - \widehat{X_{s}^{j}} \delta_{i1}  \Big] dN_{s}
\end{multline*}
Thus,
\begin{multline}
d\widehat{X_{s}^{i}X_{s}^{j}} = - \big[(B(V+M))_{ji} +(B(V+M))_{ij} - \delta_{ij}  \big] ds \\
- \big[(BV)_{1j} \widehat{X^{i}_{s}} + (BV)_{1i} \widehat{X_{s}^{j}} - \widehat{X_{s}^{i}}\delta_{j1} - \widehat{X_{s}^{j}} \delta_{i1}  \big] dN_{s}  \label{SDEforXiXjhat}
\end{multline}
We now proceed to calculate:
\begin{align*}
dV_{s}^{ij} = d( \widehat{X_{s}^{i}X_{s}^{j}} - \widehat{X_{s}^{i}} \widehat{X_{s}^{j}})
\end{align*}
Noting that 
\begin{align*}
d\widehat{X_{s}^{i}} = - (B \widehat{X_{s}})_{i} ds - ((BV)_{1i} - \delta_{i1}) dN_{s}
\end{align*}
we see that:
\begin{multline*}
d(\widehat{X_{s}^{i}}\widehat{X_{s}^{j}}) = \big[- \widehat{X_{s}^{i}}(B \widehat{X_{s}})_{j} - \widehat{X_{s}^{j}}(B \widehat{X_{s}})_{i} + ((BV)_{1i} - \delta_{i1})((BV)_{1j} - \delta_{j1}) \big] ds \\
- \big[\widehat{X_{s}^{i}}((BV)_{1j}- \delta_{j1}) + \widehat{X_{s}^{j}}((BV)_{1i} - \delta_{i1})  \big] dN_{s}
\end{multline*}
Putting this together with (\ref{SDEforXiXjhat}) gives:
\begin{multline*}
dV_{s}^{ij}= - \big[(B(V+M))_{ji} +(B(V+M))_{ij} -\delta_{ij} - \widehat{X_{s}^{i}} (B\widehat{X_{s}})_{j} - \widehat{X_{s}^{j}} (B\widehat{X_{s}})_{i} \\
+  ((BV)_{1i} - \delta_{i1})((BV)_{1j} - \delta_{j1})  \big] ds
\end{multline*}
But, we note that:
\begin{align*}
\widehat{X_{s}^{i}} (B\widehat{X_{s}})_{j} = \sum_{k} B_{jk} \widehat{X_{s}^{k}} \widehat{X_{s}^{i}} = (BM)_{ji}
\end{align*}
so finally, we deduce (\ref{DBFMEqnForV}) as required.

\end{document}